\documentclass[aps,pra,english,twocolumn,showpacs,preprintnumbers,amsmath,amssymb,floatfix,superscriptaddress,longbibliography,10pt]{revtex4-2}

\usepackage{float}
\usepackage[T1]{fontenc}
\usepackage[latin9]{inputenc}
\usepackage{graphicx}
\usepackage{amssymb}
\usepackage{babel}

\usepackage{comment}

\global\arraycolsep=2pt
\usepackage{amsmath}
\usepackage{amssymb}
\usepackage{graphicx}
\usepackage{textcomp}
\usepackage{calrsfs}
\usepackage{bm}
\usepackage{color}
\usepackage{geometry}
\usepackage{multirow}
\usepackage{array}
\newcommand{\ben}{\begin{equation*}}
\newcommand{\een}{\end{equation*}}
\newcommand{\bean}{\begin{eqnarray*}}
\newcommand{\eean}{\end{eqnarray*}}

\newcommand{\be}{\begin{equation}} 
\newcommand{\ee}{\end{equation}}
\newcommand{\bea}{\begin{eqnarray}}
\newcommand{\eea}{\end{eqnarray}}
\usepackage{orcidlink}

\colorlet{RED}{red}

\usepackage[version=3]{mhchem}

\usepackage{color}

\usepackage[mathscr]{euscript}

\usepackage{hyperref}
\hypersetup{
     colorlinks = true,
     citecolor  = magenta,  
     linkcolor  = magenta 
}

\makeatother
\usepackage{babel}

\begin{document}

\title{Prediction of a measurable sign change in the Casimir force using a magnetic fluid}

\author{Long Ma\,\orcidlink{0000-0003-3839-2604}\,}
\affiliation{Department of Physics, \href{https://www.usf.edu}{University of South Florida}, Tampa, FL, 33620, USA}

\author{Larissa In\'acio\,\orcidlink{0000-0001-8971-0591}\,}
\affiliation{\href{https://ensemble3.eu/}{Centre of Excellence ENSEMBLE3} Sp. z o. o., Wolczynska Str. 133, 01-919, Warsaw, Poland}

\author{Dai-Nam Le\,\orcidlink{0000-0003-0756-8742}\,}
\affiliation{Department of Physics, \href{https://www.usf.edu}{University of South Florida}, Tampa, FL, 33620, USA}
\author{Lilia M. Woods \orcidlink{0000-0002-9872-1847}\,}
\email{lmwoods@usf.edu}
\affiliation{Department of Physics, \href{https://www.usf.edu}{University of South Florida}, Tampa, FL, 33620, USA}

\author{Mathias Bostr{\"o}m \orcidlink{0000-0001-5111-4049}\,}
\email{mathias.bostrom@ensemble3.eu} \affiliation{\href{https://ensemble3.eu/}{Centre of Excellence ENSEMBLE3} Sp. z o. o., Wolczynska Str. 133, 01-919, Warsaw, Poland}
\affiliation{Chemical and Biological Systems Simulation Lab, \href{https://cent.uw.edu.pl/}{Centre of New Technologies, University of Warsaw}, Banacha 2C, 02-097 Warsaw, Poland}

\date{\today}

\begin{abstract}
{We demonstrate quantum levitation controlled by Casimir forces acting between a polystyrene surface and a Teflon-coated metallic substrate immersed in a mixture of Toluene and magnetite particles. This system experiences repulsion-attraction transitions in the Casimir interaction for distances where the effect is measurable. This Casimir trapping can be controlled by clever choices  of metallic and ferrofluid materials, which are directly linked to the emergence of the trapping effect. Thermal and quantum contributions are investigated in detail, showing how the optical and magnetic properties of the ferrofluid and other materials affect the magnitude of the trapping and its distance range of observability. }
\end{abstract}

\maketitle

\section{Introduction}

Understanding the Casimir entropic equilibria in parallel plate systems is crucial in Casimir-related quantum electrodynamics and applications, including Casimir switches  \cite{GongSprengCamachoLiberalEnghetaMunday2022PhysRevA.106.062824}, Casimir self-assembled structures \cite{munkhbat2021tunable,schmidt2023tunable,hovskova2025casimir}, and  mitigation of stiction in micro- and nano-electromechanical devices (MEMS and NEMS) \cite{bostrom2012casimir,stange2019building}. Casimir attraction-repulsion transition can be achieved in various ways, including by using microstructured geometry \cite{dou2014casimir,VictoriaJPCC2015}, phonon-assisted lattice vibration \cite{le2024phonon}, twisted angle in anisotropic plates \cite{pablo2023matbg, pablo2024anisotropy, hu2024twist}, suspended graphene in fluids \cite{phan2012temperature}, dielectric metamaterials on Mie resonance \cite{ye2018casimir}, and in Casimir-Lifshitz torques between anisotropic phosphorene sheets\,\cite{ThiyamPhysRevLett.120.131601_2018}, among others. Casimir trapping has also been shown to play a role in geophysics of water and ice equilibria\,\cite{Elbaum,Esteso4layerPCCP2020,BostromEstesoFiedlerBrevikBuhmannPerssonCarreteroParsonsCorkery2021,LuengoMarquez_IzquierdoRuiz_MacDowell2022,LiYang_2025_PhysRevB.111.075426}. 

Here, we present an alternative way to effectively control sign changes and Casimir quantum trapping by utilizing magnetic materials in clever combination with more conventional dielectric systems. The interplay between the optical and magnetic properties, together with the dimensions of solid film coatings, proves to give a viable pathway towards Casimir force attraction-repulsion transitions.  The proposed method for Casimir trapping builds on previous work, which showed that repulsive interaction between planar substrates separated by a ferrofluid gap is possible, but it did not explore its tuning capabilities  \cite{klimchitskaya2019impact,Klimchitskaya2019_Effect} The ultimate goal achieved here is to establish ways to control interactions in NEMS \,\cite{Munday2009,zhao2019stable,GongSprengCamachoLiberalEnghetaMunday2022PhysRevA.106.062824,shelden2023enhanced,MundayReview2024} via tunable magnetic Casimir effects\,\cite{ZhangNature2024}. Our focus is on layered systems where one can have measurable multiple sign changes in the force controlled by the 
magnetic permeability of an intermediate ferrofluid layer and the optical properties of the fluid and metallic layers in the system. 
The theoretical background of the proposed method lies within the Lifshitz formalism, established as a general theory for dispersion forces between planar surfaces\,\cite{Lifshitz1956,Dzya}, that linked the Casimir interaction to the dielectric properties of the involved materials. Furthermore, the frequency-dependent dielectric functions are related to the  refractive index and adsorption coefficients of the media, which  clarifies the relationship between forces and radiation processes proposed before the birth of quantum mechanics by Lebedev\,\cite{Lebedev2}. 

Following the original derivation, Ninham, Parsegian and co-workers ~\cite{NinhamParsegianWeiss1970,NinhamParsegian1970JCP} showed that the  Lifshitz theory can be simplified using semiclassical scattering theory; they also extended the description to account for the magnetic response properties of the materials ~\cite{Richmond_1971_magnetic}. The Lifshitz theory has been extensively explored\,\cite{Ser2018}, so one might have expected that fundamentally new insights should be difficult to find. However, we demonstrate new ways to control multiple repulsion-attraction transition distances in a four-layered system containing a magnetic fluid composed of a solution with dispersed magnetic nanoparticles. By tuning the properties of the ferrofluid in combination with clever selection of substrate dielectrics and thickness of thin surface coatings, the Casimir interaction experiences stable equilibrium potentially measurable in the laboratory via optical interferometry measurements, for example \cite{almasi2015force,sedmik2021next}. The ferrofluid plays a key role in this process, in which all types of quantum and thermal excitations are found to actively participate in forming the  trapping regime at sufficiently small separations where the force is large enough. The proposed method is beneficial in broadening the applications of NEMS and MEMS for sensors in liquid environments and biological media \cite{Stapf2024,Zu2020,Jauregui2010,Wongkasem2024}.

 

\section{Optical \& Magnetic Properties of the Considered Planar System}

\begin{figure}[H]
    \centering
    \includegraphics[width=0.9\linewidth]{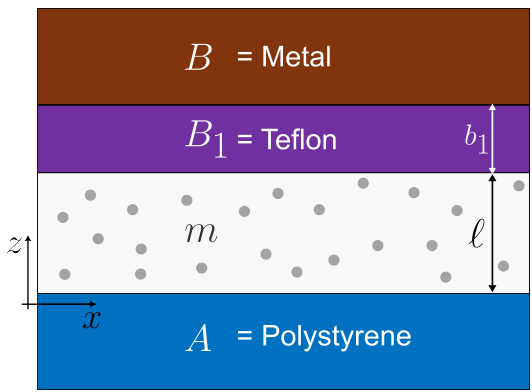}
    \caption{\label{schematicFig}Schematics of the system under consideration consisting of a ferrofluid layer (denoted as $m$) with thickness $\ell$ between a polystyrene substrate ($A$) and a  metallic substrate ($B$) covered by a Teflon layer $B_1$ of thickness $b_1$. }
\end{figure}

\begin{figure*}[htbp] 
        \centering
        \includegraphics[width=0.9\textwidth]{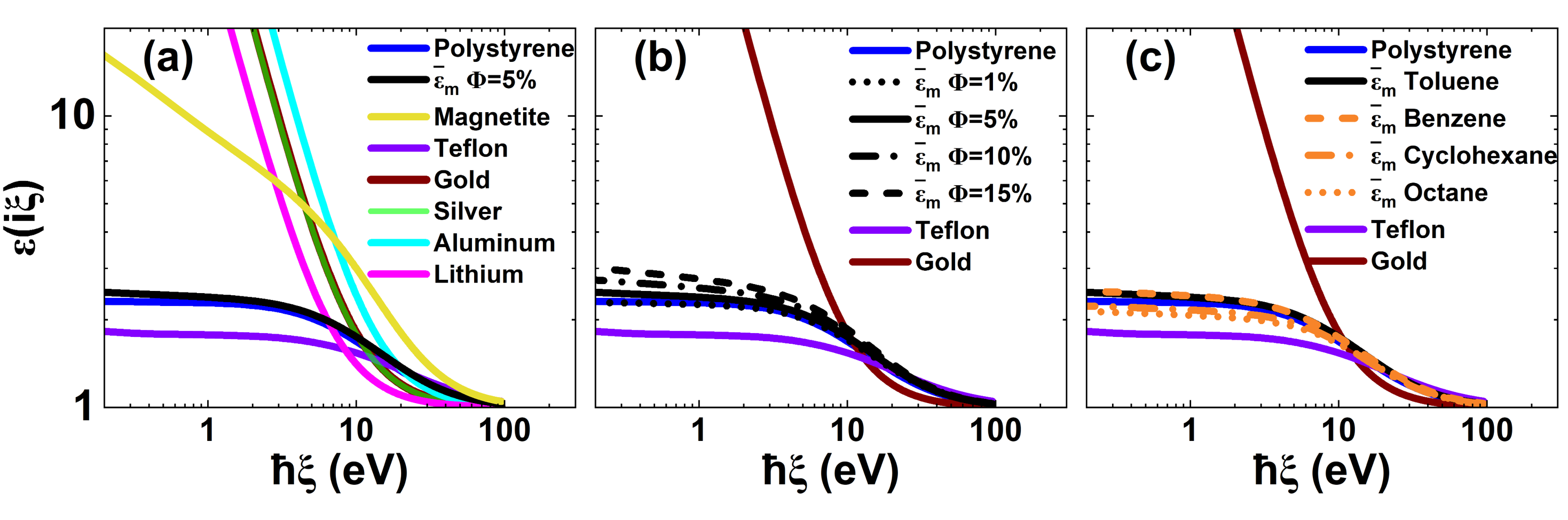} 
        \caption{Dielectric functions in the imaginary frequency range as a function of $\hbar\xi$ for the materials in the  $A-m-B_1-B$ four-layer system: (a) $A$=polystyrene, $B$=gold, silver, aluminum, lithium, $B_1$=Teflon, $m$=Toluene with $\Phi=5\%$, $D=20$ nm of Magnetite spheres; (b) $A$=polystyrene, $B$=gold, $B_1$=Teflon, $m$=Toluene, with $\Phi=1\%$, $5\%$, $10\%$, and $15\%$, $D=20$ nm of Magnetite spheres; (c) $A$=polystyrene, $B$=gold, $B_1$=Teflon, $m$=Toluene, Benzene, Cyclohexane, Octane with $\Phi=5\%$, $D=20$ nm of Magnetite spheres.
        }
        \label{fig:diel}
\end{figure*}

The layered system under consideration is given schematically in Figure \ref{schematicFig}. It consists of a planar polystyrene  substrate ($A$) and a semi-infinite metallic layer ($B$), which is coated by a Teflon film ($B_1$) of finite thickness $b_1$. The $A$ substrate and Teflon-coated metallic substrate, $B+B_1$, are separated by a gap with thickness $\ell$ filled with a ferrofluid material ($m$).  The distance-dependent Casimir interaction between the substrates in this system can be controlled in different ways, including by the choice of different metals in layer $B$, coating thickness $b_1$, and various solutions and properties of the ferrofluid in layer $m$. Since the Casimir interaction is linked to the response properties of the system, below we discuss the dielectric function and magnetic susceptibility of the involved materials. We use frequency-dependent data obtained by measuring the material's refractive index and its description of the real and imaginary parts based on the Kramer-Kronig (KK) relation with an upper integration frequency bound of 100 eV, as reported in \cite{Zwol1,zeman1987accurate,milton2004casimir}. Details of the theoretical modeling are given below in the imaginary frequency domain,  with results shown in Figure \ref{fig:diel}.

As representative examples of the types of metallic $B$ substrates, here we consider gold, silver, aluminum, and lithium. The metallic dielectric functions are modeled using the Drude model given in the imaginary frequency domain $\omega\rightarrow i\xi$ as $\varepsilon_{\rm met}(i\xi)=1+\frac{\omega_p^2}{\xi(\xi+\gamma)}$ for which the plasma frequency $\omega_p$ and relaxation frequency $\gamma$ are collected from experimental results for the considered materials \cite{milton2004casimir,zeman1987accurate}.  The dielectric functions for these metals are shown in Figure \ref{fig:diel}(a).

We also consider several solvent compositions of the ferrofluid, including Toluene, Benzene, Cyclohexane, and Octane.
Their dielectric functions are parametrized using the model {$\varepsilon_s(i\xi)=1+\sum_{i}\frac{C_i}{1+(\xi/\omega_i)2}$} with fitted oscillator strengths $C_i$ and  corresponding characteristic frequencies $\omega_i$ covering all primary absorption peaks in the infrared and ultraviolet ranges (data taken from Table 1 in \cite{Zwol1}). Furthermore, the static dielectric constants $\varepsilon_0$ of the solvents are obtained from experimental data measuring the refractive index $n_0$ based on the relation $n_0^2\approx \varepsilon_0$ for negligible infrared absorption \cite{Zwol1}.

 The ferrofluid solution contains randomly dispersed magnetic nanoparticles, here taken to be made of Fe$_3$O$_4$ whose dielectric function is also shown in Figure \ref{fig:diel}(a) using available experimental data \cite{klimchitskaya2019impact}. The effective dielectric function of the ferrofluid $\tilde\varepsilon_{m}(i\xi)$, composed of the solvent and the magnetite  nanoparticles, is calculated based on the Rayleigh mixing model \cite{sihvola1999electromagnetic}. 
\begin{align}
        \frac{\tilde\varepsilon_{m}(i\xi)-\varepsilon_{s}(i\xi)}{\tilde\varepsilon_{m}(i\xi)+2\varepsilon_{s}(i\xi)}=\Phi\frac{\varepsilon_{ \rm mag}(i\xi)-\varepsilon_{s}(i\xi)}{\varepsilon_{\rm mag}(i\xi)+2\varepsilon_{s}(i\xi)}
        \label{dielm}
\end{align}
\noindent where $\varepsilon_{\rm mag}(i\xi)$ and $\varepsilon_{s}(i\xi)$ are the dielectric functions of the magnetite and solvent, respectively, while $\Phi$ is the volume fraction of the dispersed particles. From Equation \eqref{dielm}, we find that the effective dielectric function, $\tilde\varepsilon_m(i\xi)=\varepsilon_s(i\xi) \times \frac{1+2\,\chi\,\Phi}{1-\chi\,\Phi}$ where $\chi=\frac{\varepsilon_{\rm  mag}(i\xi)-\varepsilon_{s}(i\xi)}{\varepsilon_{\rm mag}(i\xi)+2\varepsilon_{s}(i\xi)}$ has a rational dependence upon the volume fraction $\Phi$ and the $\chi$ factor. Figure \ref{fig:diel}(b) shows that in the low frequency regime, $\tilde\varepsilon_m(i\xi)$ increases as $\Phi$ is increased. Also,  $\tilde\varepsilon_m(i\xi)$ can be enhanced by taking solvents with larger $\varepsilon_s$ in the low-frequency range, as shown in Figure \ref{fig:diel}(c).

Figure \ref{fig:diel} shows expected overall behavior, as $\varepsilon_{\rm met}(i\xi)$ of all metals exhibits a divergence as $\xi\rightarrow 0$. On the other hand, the dielectric functions for the other dielectric materials have finite values consistent with $\varepsilon_0$ from experiments \cite{Zwol1}. Nevertheless, the relative magnitude of the different layers in the four-layered system in Figure \ref{schematicFig} in the low-frequency regime is crucial for attraction-repulsion transitions. The different methods of property modulations discussed here show that Casimir trapping can be achieved in many ways.

In addition to the dielectric response properties of the materials, the magnetic susceptibility of magnetite must also be considered in the Casimir interaction. Based on the distance range of interest, the model for the magnetic susceptibility is different from 1 for only the zero-frequency term. The static (zero-frequency) effective permeability for the ferrofluid can be modeled as
\,\cite{klimchitskaya2019impact,Klimchitskaya2019_Effect},
\begin{equation}
    \tilde\mu_m(0)=1+\frac{2\pi^2\Phi}{9}\frac{M_s^2 D^3}{k_{B} T}.
    \label{mu}
\end{equation}
where  $M_s$ is the saturation magnetization per unit volume and $D$ is the nanoparticle average diameter. Also, $k_B$ is the Boltzmann constant and $T$ is the temperature. The above expression suggests that different material characteristics can be used to modulate $\tilde\mu_m(0)$, which ultimately can modify the Casimir pressure as shown in what follows.

\section{Casimir interaction in planar polystyrene-ferrofluid-Teflon-metal systems}

 The Casimir pressure between the Teflon-coated metallic and polystyrene substrates across the ferrofluid filled gap can be calculated using the Lifshitz formalism given in Matsubara frequencies $\xi_n=\frac{2\pi n k_B T}{\hbar}$ 
 \cite{Lifshitz1956,Dzya},
 \begin{widetext}
\begin{equation}
    P=-k_BT\sum_{n=0}^{\infty}{}^\prime \int_{0}^{+\infty}   \frac{d^2\textbf{k}_{\parallel}}{(2\pi)^2}\sum_{\alpha}\frac{2 k_{z{\rm m}}\, R_{\rm Am}^{\alpha}(i\xi_n)R_{\rm eff}^{\alpha}(i\xi_n)e^{-2k_{z{\rm m}}\ell}}{1-R_{\rm Am}^{\alpha}(i\xi_n)R_{\rm eff}^{\alpha}(i\xi_n)e^{-2k_{z{\rm m}}\ell}}  ,
    \label{P}
\end{equation}
\end{widetext}
where $\ell$ is the separation distance in the nanocavity, as shown in Figure \ref{schematicFig}. The prime in the sum indicates there is 1/2 weight of the first term ($n=0$). Also, $k_{z{\rm m}}=\sqrt{k_{\parallel}^2+{\tilde\varepsilon_m\tilde\mu_m\xi_n^2}/{c^2}}$ is the wavevector in the z-direction across the  ferrofluid and $k_{\parallel}^2=k_x^2+k_y^2$. The Fresnel reflection coefficients 
$R_{\rm Am}^\alpha$ corresponds to electromagnetic scattering at the interface of layers $A$ and $m$, where the superscript $\alpha$=TE (transverse electric) or TM (transverse magnetic) refers to the polarization modes. 

The Fresnel reflection coefficients $R_{\rm eff}^{\alpha}(i\xi_n)$ capture TE and TM scattering from the metallic $B$ substrate coated by the Teflon film $B_1$. These are effective coefficients obtained via a multi-scattering formalism
\cite{parsegian2005van,podgornik2003reformulation}, 
\begin{equation}
    R_{\rm eff}^{\alpha}(i\xi_n)=\frac{R_{{\rm B}_1{\rm m}}^{\alpha}(i\xi_n)+R_{{\rm BB}_1}^{\alpha}(i\xi_n)e^{-2k_{z{\rm B}_1}b_1}}{1+R_{{\rm B}_1{\rm m}}^{\alpha}(i\xi_n) R_{{\rm BB}_1}^{\alpha}(i\xi_n) e^{-2k_{z{\rm B}_1}b_1}}
    \label{Reff}
\end{equation}
where $k_{z{\rm B}_1}=\sqrt{k_{\parallel}^2+{\varepsilon_{B_1}(i\xi_n)\xi_n^2}/{c^2}}$ is the wavevector in the z-direction in the Teflon layer, and $b_1$ is its  thickness. The Fresnel coefficients $R_{{\rm BB}_1}^{\alpha}(i\xi_n)$ and $R_{{\rm B}_1{\rm m}}^{\alpha}(i\xi_n)$ refer to the scattering at the $B/B_1$ and $B_1/m$ interfaces, respectively. By solving the Maxwell's equations under standard boundary conditions, we find 
\begin{eqnarray}
        R_{ij}^{\rm TM}(i\xi_n)&=&\frac{k_{zj} \, \varepsilon_i(i\xi_n)-k_{zi}\, \varepsilon_j(i\xi_n)}{k_{zj}\, \varepsilon_i(i\xi_n)+k_{zi}\, \varepsilon_j(i\xi_n)} ,
        \label{RTM} \\
        R_{ij}^{\rm TE}(i\xi_n)&=&\frac{k_{zj}\, \mu_i(i\xi_n)-k_{zi}\, \mu_j(i\xi_n)}{k_{zj}\, \mu_i(i\xi_n)+k_{zi}\, \mu_j(i\xi_n)} ,
        \label{RTE}
\end{eqnarray}
for the $i/j$ ($i,j=A,\space B,\space B_1,\space m $) interfaces with $k_{zi}=\sqrt{k_{\parallel}^2+{\varepsilon_{i}\mu_{i}\xi_n^2}/{c^2}}$ being the $z$-component of the wave vector in medium  $i$. 

\section{Results and Discussion}

\begin{figure*}[htbp] 
        \centering
        \includegraphics[width=\textwidth]{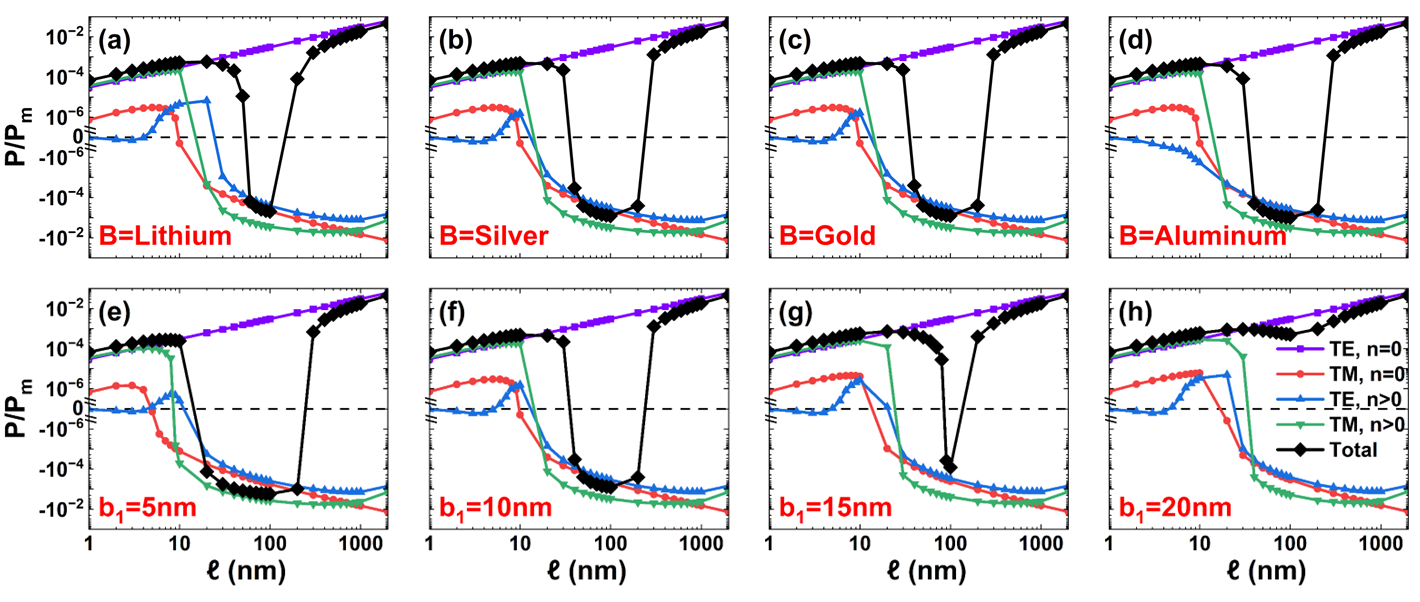} 
        \caption{Casimir pressures normalized to the perfect conductor limit $P_m(\ell)=-\frac{\pi^2\hbar c}{240\ell^4}$ in the four-layer system from Figure \ref{schematicFig}, where the ferrofluid is Toluene with magnetite nanoparticles with average diameter $D=20$ nm and volume fractions $\Phi$=5\%. For panels (a)-(d), the thickness of the Teflon layer is $b_1=10$ nm while the metallic material B varies in plasma frequency $\omega_P$ and relaxing frequency $\gamma$,  such as (a) lithium ($\hbar\omega_P=6.45$ eV, $\gamma=0.13$ eV), (b) silver ($\hbar\omega_P=8.9$ eV, $\gamma=0.02$ eV), (c) gold ($\hbar\omega_P=9$ eV, $\gamma=0.03$ eV), and (d) aluminum ($\hbar\omega_P=12.04$ eV, $\gamma=0.13$ eV) \cite{zeman1987accurate, milton2004casimir}. For panels (e)-(h), the metallic layer is gold while the thickness of the Teflon layer varies, such as (e) $b_1=5$ nm, (f) $b_1=10$ nm, (g) $b_1=15$ nm, (h) $b_1=20$ nm.  The different contributions from TE and TM modes with zero-frequency ($n=0$) and summed  frequencies ($n>0$) terms are also displayed.}
        \label{fig:3BB1}
\end{figure*}

The theoretical framework described in the previous section can now be used to calculate the Casimir pressure between the planar materials separated by the ferrofluid, as shown in Figure \ref{schematicFig}. 
We present a comprehensive understanding of the Casimir interaction and its various contributing factors by investigating the role of the specific metallic and solvent materials and the various structural characteristics of the system.

We first probe into the role of the specific metal of layer $B$. In metallic systems, the Casimir force can be tailored by modulating the plasma frequency of the materials. The optimization of the plasma frequency is important in the design of epsilon-near-zero optical materials  \cite{camacho2022engineering} and for enhancing near-field radiation \cite{ma2025radiativeheattransfer2d} among other applications. In the context of Casimir interactions, surface plasmons and surface plasmon polaritons have also been instrumental in finding effective pathways of control \cite{camacho2022engineering, gong2023effect}.  The considered system in Figure \ref{schematicFig} can be constructed with different metallic substrates, and the calculated Casimir pressure is shown for several materials in  Figure \ref{fig:3BB1}(a-d).  The dielectric response for the metals is taken via the Drude model, as discussed previously, with parameters obtained from available experimental data (see Figure \ref{fig:3BB1}(a-d)).

Figure \ref{fig:3BB1}(a-d) shows that the total Casimir pressure exhibits similar behavior for all metals with a trapping region in the submicron range. To further understand the contributions from the different types of electromagnetic polarization, the Casimir pressure is decomposed into TE and TM modes, explicitly showing the thermal zero Matsubara frequency term and the summation of all Matsubara frequencies with $n>0$. 

We note that the thermal TE and TM contributions hardly change for the different metals. This is verified by the asymptotic behavior of the Casimir pressure obtained in terms of effective permeabilities and dielectric responses (see Equation \eqref{eqn:te-thermal} and \eqref{eqn:tm-thermal} in Appendix) This is also reflected in the Fresnel coefficients for the metal/Teflon interface for large and small distances since the zero-frequency limit gives  $R^{\rm TE}_{BB_{1}}(0) =0$ and $R^{\rm TM}_{BB_{1}}(0) =1$, and for the large frequency limit  $\lim\limits_{\xi\to\infty} R^{\rm TE}_{BB_{1}}( i\xi_n)=0, \lim\limits_{\xi\to\infty} R^{\rm TM}_{BB_{1}}( i\xi_n)=0$ for both polarizations. The material dependence upon the different metals is mostly seen in the $\ell=(10,50)$ nm window for the  TE mode with $n>0$, where the plasma frequency controls the attraction-repulsion transition. 
A larger plasma frequency increases the repulsive part from the TE $n>0$ contribution, which is the main factor in modulating the Casimir trapping region for this system.

\begin{figure*}[htbp] 
        \centering
        \includegraphics[width=\textwidth]{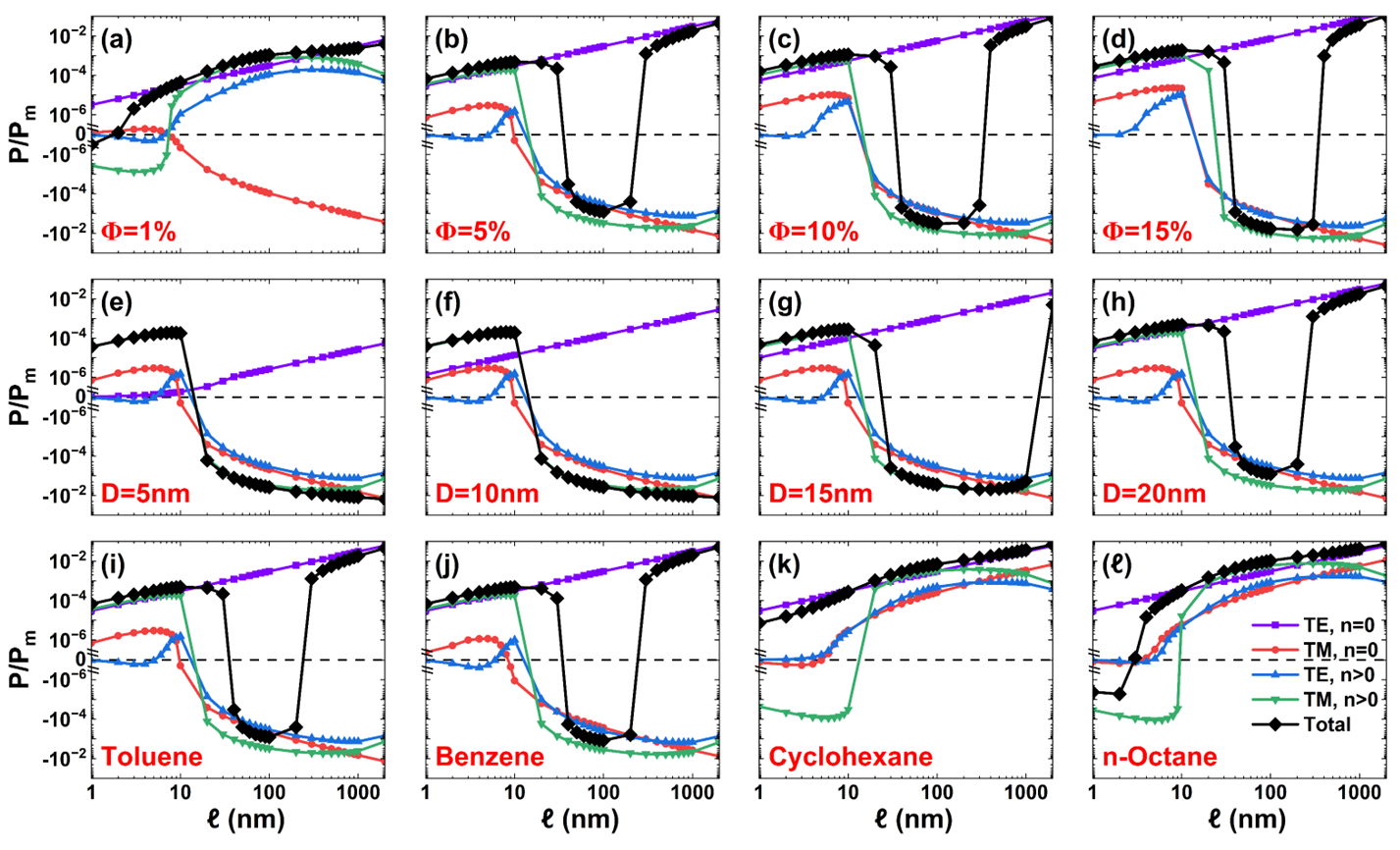} 
        \caption{Casimir pressures normalized to the perfect conductor limit $P_m(\ell)=-\frac{\pi^2\hbar c}{240\ell^4}$ for the system in Figure \ref{schematicFig}. For sub-panels (a)-(d), the volume fraction of the magnetite particles with $D=20$ nm in the Toluene ferrofluid are (a) $\Phi$=1\%, (b) $\Phi$=5\%, (c) $\Phi$=10\%, and (d) $\Phi$=15\%. For sub-panel (e)-(h), the nanoparticles are dispersed in Toluene with a volume fraction $\Phi=5\%$ with diameters taken as (e) $D=5$ nm, (f) $D=10$ nm, (g) $D=15$ nm, and (h) $D=20$ nm. For sub-panels (i)-($l$),  nanoparticles with $D=20$ nm and volume fraction $\Phi=5\%$ are dispersed in (i) Toluene, (j) Benzene, (k) Cyclohexane, and ($\ell$) Octane. In all cases, $B$=gold and the thickness of the Teflon layer is $b_1=10$ nm.
        The different contributions from TE and TM modes with zero-frequency (n=0) and summed non-zero frequencies ($n>0$) terms are also given with distinct line symbols and line styles.}
        \label{fig:4m}
\end{figure*}

Another factor that can affect the Casimir interaction is the thickness of the Teflon layer. In Figure \ref{fig:3BB1}(e-f), numerical results are shown for the pressure $P$ between the polystyrene layer covered by Teflon and gold with Ferrite ferrofluid as a function of the gap separation $\ell$ for different thicknesses of the Teflon layer. We observe that a stable trapping range is possible. As $b_1$ increases, the repulsive range shrinks and for $b_1=20$ nm, the Casimir pressure exhibits no repulsion and it becomes very close to the one for the thermal limit from the TE $n=0$ contribution with the scaling law $P \sim\ell^{-3}$ \cite{le2022dispersive}. This scaling law and the $b_1$ independence is consistent with the analytical expression of the Casimir pressure for TE n=0 (see Equation \eqref{eqn:te-thermal} in Appendix).

To understand how the Teflon thickness $b_1$ affects the interaction, we focus on the characteristic behavior of the Fresnel reflection coefficients $R_{BB1}^\alpha(i\xi_n)$ and $R_{{\rm B}_1{\rm m}}^\alpha(i\xi_n)$. From Equations \eqref{RTE}, we find that 
$R_{{\rm BB}_1}^{\rm TE}(0)=0$
and  
$R_{{\rm B}_1{\rm m}}^{\rm TE}(0)=\frac{\tilde\mu_m(0)-1}{\tilde\mu_m(0)+1}$, which means that 
$R_{\rm eff}^{\rm TE}(0)=R_{{\rm B}_1{\rm m}}^{\rm TE}(0)$. This is consistent with the finding that the thermal TE contribution is unaffected by the Teflon thickness (See Equation \eqref{eqn:te-thermal} in Appendix). On the other hand, {\footnotesize $R_{{\rm BB}_1}^{\rm TM}(0)=[\,\varepsilon_{B_1}(0)-\varepsilon_{B}(0)\,]\,/\,[\,\varepsilon_{B_1}(0)+\varepsilon_{B}(0)\,]$ so that {\footnotesize $R_{\rm eff}^{\rm TM}(0)\approx [\,R_{{\rm BB}_1}^{\rm TM}(0) + e^{-2k_{\parallel}b_1}\,]\,/\,[\,1+R_{{\rm BB}_1}^{\rm TM}(0) e^{-2k_{\parallel}b_1}\,]$}} shows that the attraction-repulsion transition of the thermal TM contribution can be shifted by changing the thickness $b_1$. The particular $b_1$ dependence of the thermal TM contribution is given in Equation \eqref{eqn:tm-thermal-large}. The scaling law is $\ell^{-3}$ for both small $\ell \ll b_1$ and large limits $\ell \gg b_1$, but there is a sign change when going from $\ell \ll b_1$ to $\ell \gg b_1$, consistent with the numerical results shown in Figure \ref{fig:3BB1}.
 


From Equations \eqref{Reff} we also find that the TE and TM  contributions with summed $n>0$ Matsubara frequencies are 
thickness-dependent in the $\ell=(10,100)$ nm range.  For $\ell\gg b_1$, the effective Fresnel coefficient {\footnotesize $R_{\rm eff}^{\alpha}(i\xi_n)\approx R_{{\rm B}_1{\rm m}}^{\alpha}(i\xi_n)$}, and the Casimir pressure is not affected by $b_1$.  For  $\ell\ll b_1$, the effective coefficients are also independent of $b_1$ since 
{\footnotesize $\lim\limits_{\xi\to\infty}R_{\rm eff}^{\rm TE}(i\xi_n)=\lim\limits_{\xi\to\infty}R_{\rm eff}^{\rm TM}(i\xi_n)=0$}. 
Therefore, the Teflon thickness plays a role in the intermediate $b_1\sim \ell=(10,100)$ nm region, specifically around the attraction-repulsion transition point, as illustrated in Figure \ref{fig:3BB1}.

This analysis shows that the Teflon layer thickness is mostly effective when $b_1\sim\ell$ ($10\sim100$ nm in our case), with changes coming mostly from the TE $n>0$ frequencies contribution. Ultimately, the overall Casimir pressure is determined by the competition between the attractive TE thermal terms and the combined effect of repulsive thermal (TM $n=0$) and quantum (TE and TM $n>0$) contributions. Increasing  $b_1$ reduces the magnitude of the repulsive part of the pressure in the $\ell\approx 20\sim 50$ nm range, leading to the disappearance of the Casimir trapping.

The Casimir interaction in the layered system from Figure \ref{schematicFig} can also be modulated by changing the properties of the ferrofluid. Here we investigate the role of the nanoparticle volume fraction $\Phi$, their average diameter $D$, and the type of solvent with results summarized in Figure \ref{fig:4m}. Overall, we see that the trapping occurs for larger volume fractions (Figure 4(a-b)), but this effect may disappear for larger diameters (Figure 4(e-g)) due to the competition of different attractive and repulsive contributions for
the Casimir pressure. The solvent substance also plays a role in the attraction-repulsion transitions for the total pressure.

As evident from Equations \eqref{dielm} and \ref{mu}, the nanoparticle volume fraction affects both the effective dielectric and magnetic properties of the ferrofluid following a positive correlation. Figure \ref{fig:4m}(a-d) shows the zero-frequency TE contribution is always attractive following the characteristic $\ell^{-3}$ scaling law, however, its magnitude depends on the volume fraction following $\sim\Phi^2$ behavior (See Equation \eqref{eqn:te-thermal} in Appendix). The zero-frequency TM contribution exhibits attractive and repulsive regions, which can also be controlled by $\Phi$.  Making $\Phi$ larger increases the dielectric function of the ferrofluid $\tilde\varepsilon_m$ as shown in Figure \ref{fig:diel}, thus the pressure magnitude of the thermal TM contribution, which is determined by Fresnel coefficients $R_{\rm eff}^{\rm TM}(0)$ and $R_{\rm Am}^{\rm TM}(0)$, is also enhanced (See Equations \eqref{eqn:tm-thermal}--\eqref{eqn:tm-thermal-small} in Appendix).

Figure \ref{fig:4m}(a-d) also shows that the TM and TE terms with summed $n>0$ frequencies also experience significant transformations as $\Phi$ increases. For $\Phi=1 \%$, we see that $\tilde\varepsilon_m\sim\varepsilon_s$ (see Equation \eqref{dielm}). As discussed previously, for $b_1\sim\ell$ the Fresnel coefficients {\footnotesize $R_{\rm Am}^{\rm TM}(i\xi_n)$} and {\footnotesize $R_{\rm eff}^{\rm TM}(i\xi_n)$} have opposite signs in the small separation limit, so the finite frequency TM term is repulsive for smaller $\ell$. This repulsion is further traced to the $\varepsilon_{\rm A} >\tilde\varepsilon_{\rm m}>\varepsilon_{B_1}$ ordering of the dielectric functions (see Figure \ref{fig:diel}(b) for $\tilde\varepsilon_m$ $\Phi=1\%$). As the distance $\ell$ increased, the role of $b_1$ diminishes and the Casimir pressure for the $n>0$ TM modes is mostly determined by the optical response of the polystyrene substrate and the ferrofluid. Thus the Casimir interaction due to the $n>0$ TM contribution becomes attractive at larger distances $\ell$ (see Figure \ref{fig:4m}(a). A similar behavior is also observed for the TE $n>0$ contribution for $\Phi=1\%$, where the attraction-repulsion transition occurs at separations $\ell$ that are very similar to those for the TM $n>0$ modes. 

For larger volume fractions, from $\Phi=5\%$ and above, the $n>0$ TM term reverses this trend, sustaining a strong repulsive range for $\ell>10$ nm also linked to the $\tilde\varepsilon_{\rm m}>\varepsilon_{\rm A}>\varepsilon_{B_1}$ ordering of the dielectric functions. A similar behavior is observed for the TE $n>0$ contribution. Ultimately, as $\Phi$ is increased, the role of $\tilde\mu_m$ grows, and the interplay between the attractive zero-frequency TE mode and repulsive TE $n>0$ modes, together with all TM modes, results in a stable equilibrium region $\ell\sim (30,300)$ nm. Note, however, since both the effective dielectric response of the ferrofluid experiences positive correlation with $\Phi$, the attractive pressure from the zero-frequency TE term and the repulsive pressure from all TM and non-zero-frequency TM modes are enhanced simultaneously. Thus, further increasing $\Phi$ does not result in significant changes in this stable equilibrium region.

In Figure \ref{fig:4m}(e-h), we also show the effect of the average nanoparticle diameter on the Casimir pressure. The role of $D$ comes mainly through $\tilde\mu_m(0)$ primarily affecting the thermal TE contribution, whose magnitude increases significantly as the nanoparticle size increases (See Equation \eqref{eqn:te-thermal} in Appendix). This is unlike the effect of $\Phi$ whose role is not only felt in the TE $n=0$ mode but also in all other contributions, as discussed previously. Figure \ref{fig:4m}(e-h) shows that other terms, except for the TE $n=0$ contribution, have a similar behavior showing attraction-repulsion transition at $\ell\sim20$ nm, and they are largely unaffected by $D$. As a result, a stable equilibrium region whose range along the separation axis is affected by the diameter of the nanoparticles becomes possible, as shown for $D=15$ and $D=20$ nm (Figure \ref{fig:4m}(g-h)). 

The particular ferrofluid solvent is another component that can significantly affect the Casimir interaction.  In our system, the optical properties of the solvent not only determine the dielectric response of the ferrofluid dispersion but also generate effective reflection coefficient and net Casimir pressure between the two substrates. In Figure \ref{fig:4m}(i-l), we show the results of the Casimir pressure for different solvents, including Toluene, Benzene, Cyclohexane, and n-Octane. We find that Toluene and Benzene, with commensurate dielectric response functions (see Figure \ref{fig:diel}), generate similar  behavior for the pressure with a defined trapping zone in the $\ell\sim(10,100)$ nm range. However, Cyclohexane and Octane whose dielectric functions are smaller than the ones for Toluene and Benzene, primarily create attractive Casimir interaction at the sub-micron range. The opposite behaviour observed in two scenarios (Figure \ref{fig:4m} (i-j) and (k-l)) is consistent with the prediction from analytical expressions \eqref{eqn:tm-thermal-large}--\eqref{eqn:tm-thermal-small} in the Appendix. Particularly, $\varepsilon_{B_1} (0) < \varepsilon_A (0) < \tilde{\varepsilon}_m(0)$ holds for Toluene and Benzene with concentration larger than $2\%$ of dispersed particles and causes the attractive-to-repulsive transition when increasing the separation gap $\ell$. Conversely, $\varepsilon_{B_1} (0) < \tilde{\varepsilon}_m(0) < \varepsilon_A (0) $ holds for Cyclohexane and n-Octane with concentration less than $10 \%$ of nanoparticles, resulting in the repulsive-to-attractive transition versus separation gap $\ell$. The decomposition of the polarization mode further indicates the sign flips in both TM $n\geq0$ and TE $n>0$ modes between the two scenarios (Figure \ref{fig:4m} (i-j) and (k-l)), while the TE $n=0$ terms are rarely changed. 

Figures \ref{fig:3BB1} and \ref{fig:4m} show that a stable equilibrium originating entirely from the Casimir interaction can be achieved in several ways in the multilayered system we have designed. Although the interplay between the dielectric and magnetic properties of the involved layers is complex with several tunable factors, the achieved trapping is in the $10-200$ nm range with magnitude of the repulsion reaching $P\sim10^{-7}$ mPa. This falls within the range of experimental observations. In particular, the Casimir and NonNewtonian force EXperiment (CANNEX) based on optical interferometry reported in \cite{almasi2015force, sedmik2021next}) have shown that the Casimir force in the $10^{-8}\sim 10^{-4}$ mPa range for sub-micron separations are indeed achievable in the laboratory. 

\section{Conclusion}

In summary, we have calculated the Casimir interaction in a novel multi-layered system consisting of a polystyrene substrate and Teflon-coated metallic layer separated by a gap filled with a ferrofluid. 
It is found that the sign of the Casimir force can be changed in many ways by tuning various parameters of the dielectric and magnetic properties of the involved materials. From the theoretical analysis within the Lifshitz formalism, we show the direct property-interaction relationship helping us understand the underlying factors of interaction control. In particular, changing the type of metal and thickness of the Teflon layer are most effective for changes in the TE polarization contributing to the quantum mechanical Casimir force in the submicron distance range. The type of solvent in the ferrofluid, however, influences not only the quantum mechanical TE contribution but also the thermal and quantum mechanical TM modes. On the other hand, the volume fraction is linked to changes in both thermal and quantum contributions (including TE and TM modes), while the diameter of the nanoparticle spheres in the ferrofluid changes the quantum trapping only by affecting the thermal TE modes.

Regardless of the chosen method of control, our calculations show that a stable regime of quantum trapping in the submicron regime with measurable magnitude is achievable and experimentally measurable. The diversity of different ways to modulate the system is also advantageous from an experimental point of view, as one can accommodate various laboratory capabilities that can yield measurable results of Casimir trapping \cite{almasi2015force, sedmik2021next}. Our study gives useful guidelines for Casimir force control, which is useful for the development of novel micro- and nanomechanical devices. 
 
\begin{acknowledgments}
L.I. and M.B. part of the research is part of the project No. 2022/47/P/ST3/01236 co-funded by the National Science Centre and the European Union's Horizon 2020 research and innovation programme under the Marie Sk{\l}odowska-Curie grant agreement No. 945339. 
 Institutional and infrastructural support for the ENSEMBLE3 Centre of Excellence was provided through the ENSEMBLE3 project (MAB/2020/14) delivered within the Foundation for Polish Science International Research Agenda Programme and co-financed by the European Regional Development Fund and the Horizon 2020 Teaming for Excellence initiative (Grant Agreement No. 857543), as well as the Ministry of Education and Science initiative "Support for Centres of Excellence in Poland under Horizon 2020" (MEiN/2023/DIR/3797). L.M.W. acknowledges financial support from the US Department
of Energy under Grant No. DE-FG02-06ER46297.   
 \end{acknowledgments}
 
\appendix*

\section*{Appendix: The thermal contributions to the Casimir pressure}

The $n = 0$ TE term is found directly from Equation \eqref{P} as
\begin{eqnarray}
    \label{eqn:te-thermal}
    P_{n=0}^{TE} (\ell) && = - \frac{k_B T}{8 \pi \ell^3} \text{Li}_3 \left[ \left( \dfrac{\tilde{\mu}_m(0) - 1}{\tilde{\mu}_m(0) + 1} \right)^2 \right]\nonumber\\
\end{eqnarray}
This term is always attractive and it scales as $\ell^{-3}$. Given Equation \eqref{mu} in the main text, $P_{n=0}^{TE}$ is tuned by the concentration $\Phi$ and diameter $D$ of the nanoparticles in the ferrofluid. When $\frac{\pi^2 \Phi M_S^2 D^3}{9 k_B T} \ll 1$, Equation \eqref{eqn:te-thermal} becomes $ {P_{n=0}^{TE} (\ell)} \approx - \frac{k_B T}{8 \pi \ell^3}  \left(\frac{\pi^2 \Phi M_S^2 D^3}{9 k_B T} \right)^2 \propto \Phi^2 D^6$.

The $n = 0$ TM term is also found from Equation \eqref{P},
\begin{eqnarray}
    \label{eqn:tm-thermal}
    && P_{n=0}^{TM} (\ell) \approx - \frac{k_B T}{8 \pi \ell^3} \text{Li}_3 \left[ \mathcal{R}\left( e^{-2b_1/\ell} \right) \right], \\
    && \mathcal{R}\left( x \right) = \dfrac{ \left(\frac{\varepsilon_{A}(0) - \tilde{\varepsilon}_m(0)}{\varepsilon_{A}(0) + \tilde{\varepsilon}_m(0)}\right) \left[\left(\frac{\varepsilon_{B_1}(0) - \tilde{\varepsilon}_m(0)}{\varepsilon_{B_1}(0) + \tilde{\varepsilon}_m(0)}\right) + x \right] }{1+ \left(\frac{\varepsilon_{B_1}(0) - \tilde{\varepsilon}_m(0)}{\varepsilon_{B_1}(0) + \tilde{\varepsilon}_m(0)}\right) x} \nonumber .
\end{eqnarray}
In the case of $\ell \gg b_1$, we find that
\begin{eqnarray}
    \label{eqn:tm-thermal-large}
   && P_{n=0}^{TM} (\ell)  \approx - \frac{k_B T}{8 \pi \ell^3} \left\{ \text{Li}_3 \left(\frac{\varepsilon_{A}(0) - \tilde{\varepsilon}_m(0)}{\varepsilon_{A}(0) + \tilde{\varepsilon}_m(0)}\right) \right. \nonumber\\
    && \left.- \frac{2 \tilde{\varepsilon}_m(0)}{\varepsilon_{B_1}(0)}\text{Li}_2 \left(\frac{\varepsilon_{A}(0) - \tilde{\varepsilon}_m(0)}{\varepsilon_{A}(0) + \tilde{\varepsilon}_m(0)}\right) \frac{b_1}{\ell} \right\}.
\end{eqnarray}
Thus the Casimir interaction is independent of the thickness $b_1$.

In the case of $\ell \ll b_1$, the Casimir interaction is independent of the metallic layer thickness $B$ as found asymptotically,
\begin{eqnarray}
    \label{eqn:tm-thermal-small}
    && P_{n=0}^{TM} (\ell) \approx  - \frac{k_B T}{8 \pi \ell^3} \times \nonumber\\
    && \times \text{Li}_3 \left[\left(\frac{\varepsilon_{A}(0) - \tilde{\varepsilon}_m(0)}{\varepsilon_{A}(0) + \tilde{\varepsilon}_m(0)}\right)  \left(\frac{\varepsilon_{B_1}(0) - \tilde{\varepsilon}_m(0)}{\varepsilon_{B_1}(0) + \tilde{\varepsilon}_m(0)}\right) \right]. \nonumber\\
\end{eqnarray}
From these limits, we can show that by tuning the ferrofluid 's static dielectric constant $\tilde{\varepsilon}_m(0)$, one can manipulate the sign of thermal TM Casimir pressure in small and large seperation distance $\ell$. In particular, since $\varepsilon_{B_1} (0) = 2.1$ while $\varepsilon_A (0) = 2.4$, different situations may occur, such as 
\begin{itemize}
   \item $\varepsilon_{B_1} (0) < \varepsilon_A (0) < \tilde{\varepsilon}_m(0)$, then for $\ell \ll b_1$ the $n=0$ TM contribution to the force is attractive and for $\ell \gg b_1$ the force is repulsive;
   \item $\varepsilon_{B_1} (0) < \tilde{\varepsilon}_m(0) < \varepsilon_A (0) $, then the $n=0$ TM contribution is repulsive for  $\ell \ll b_1$, while it is attractive for $\ell \gg b_1$;
   \item $\tilde{\varepsilon}_m(0) < \varepsilon_{B_1} (0)  < \varepsilon_A (0) $, the $n=0$ TM contribution is always attractive for any separation distance $\ell$.
\end{itemize}

\bibliography{Casimir}


\end{document}